\documentclass[9pt,conference]{IEEEtran}
\IEEEoverridecommandlockouts

\usepackage{cite}
\usepackage{amsmath,amssymb,amsfonts}
\usepackage{algorithmic}
\usepackage{graphicx}
\usepackage{textcomp}
\usepackage{xcolor}
\usepackage{multirow}
\usepackage{siunitx} 
\usepackage{array}
\usepackage{subcaption}
\usepackage{url}
{

\def\BibTeX{{\rm B\kern-.05em{\sc i\kern-.025em b}\kern-.08em
    T\kern-.1667em\lower.7ex\hbox{E}\kern-.125emX}}
\begin{document}

\title{Energy Consumption Trends in Sound Event Detection Systems}

\author{\IEEEauthorblockN{Constance Douwes, Romain Serizel}
\IEEEauthorblockA{
\textit{Université de Lorraine, CNRS, Inria, Loria}
\\Nancy, France}}

\maketitle

\begin{abstract}
Deep learning systems have become increasingly energy- and computation-intensive, raising concerns about their environmental impact. As organizers of the Detection and Classification of Acoustic Scenes and Events (DCASE) challenge, we recognize the importance of addressing this issue. For the past three years, we have integrated energy consumption metrics into the evaluation of sound event detection (SED) systems. In this paper, we analyze the impact of this energy criterion on the challenge results and explore the evolution of system complexity and energy consumption over the years. We highlight a shift towards more energy-efficient approaches during training without compromising performance, while the number of operations and system complexity continue to grow. Through this analysis, we hope to promote more environmentally friendly practices within the SED community.\footnote{The data related to the submissions used for the analysis will be added after the review process.}
\end{abstract}


\section{Introduction}

For several years now, deep learning (DL) has revolutionized the fields of signal processing and machine listening \cite{purwins2019deep}, driven by a constant quest for better performance enabled by high-performance hardware. However, as DL systems become more powerful, the associated computational overhead also increases \cite{amodei2018ai, sevilla2022compute,thompson2007computational}, resulting in high energy costs. This growing energy demand raises concerns about the associated environmental impact~\cite{henderson2020towards,gupta2021chasing,lacoste2019quantifying}, concerns that first emerged in the field of natural language processing \cite{strubell2020energy} and, more recently, within the audio processing and machine listening communities \cite{douwes2021energy,parcollet2021energy}. 

As organizers of the Detection and Classification of Acoustic Scenes and Events (DCASE) challenge, the reference challenge for sound event detection (SED) systems, we acknowledge our role in this pursuit of performance and the associated energy cost \cite{serizel2023performance}. Therefore, starting from 2022, energy and compute considerations were integrated in task 4 of the challenge \cite{ronchini2022description}. This task aims to study SED systems on a heterogeneous dataset with potentially missing labels and varying levels of detail \cite{turpault2019sound}. The introduction of this new criterion raised questions about the comparability of energy measurements under different hardware setups. For example, Ronchini et al. proposed the use of a trivial normalization for energy consumption that takes into account hardware disparities \cite{ronchini2022description}. From this starting point, we performed an in-depth analysis of different normalization strategies showing the high influence of the reference system and highlighting that normalization is far from straightforward \cite{douwes2024normalizing}. 

Another way to account for the energy consumption of neural networks-based systems is to consider their computational cost, as the number of parameters or the number of operations. However, we have shown that there is no simple correlation between energy consumption and the computational cost across different simple architecture types (e.g., fully connected, convolutional, recurrent layers) for training and testing SED systems on the same setup \cite{douwes2024computation}. This is particularly true when the hardware and the implementation are different, as in the DCASE challenge \cite{ronchini2024performance}. Nevertheless, considering all of these metrics together provides a more comprehensive view of the environmental cost of such systems~\cite{schwartz2020green}.

In this paper, we explore the evolution of the energy consumption and the computational load of the systems submitted to DCASE task 4 over the past three years. We build on last year's analysis of the 2023 submissions \cite{ronchini2024performance}, which focused on performance and energy balance, and incorporate the new 2024 entries. We do not address the link between energy consumption and computational load, as this aspect was already covered by last year's analysis and led to the above-mentioned conclusions. Instead, we focus on the key differences between the DCASE 2023 and DCASE 2024 submissions, particularly due to the availability of normalized energy consumption data. Additionally, we incorporate to our analysis novelties from this year's submissions \cite{cornell2024dcase}, including the detailed energy consumption measurements with a focus on GPU utilization, and the introduction of a new performance metric.

\section{Analysis setup and novelties}

This study is based on DCASE task 4 submissions in 2022, 2023, and 2024. We encourage the reader to visit the DCASE challenge website to learn more about system submissions\footnote{https://dcase.community/challenge2024/task-sound-event-detection-with-heterogeneous-training-dataset-and-potentially-missing-labels}. In the following, we detail the evolution of task 4, performance metrics and energy measurements over the years.

\subsection{Task setup and performance metrics}

Task 4 of the DCASE challenge has evolved over the years. In 2022, the task focused on detecting sound events using systems trained on weakly labeled data (without timestamps) and synthetic soundscapes. In 2023, task 4 was divided into two subtasks: 4a, which was a direct continuation of 2022, and 4b, which introduced soft labeled data (which value can be between 0 and 1). In 2024, task 4 unified these subtasks, with the goal of detecting sound events while leveraging training data with varying levels of annotation granularity, such as temporal resolution and soft/hard labels \cite{cornell2024dcase}. All the baselines systems are based on the same mean-teacher model \cite{turpault2019sound}, with some adjustments for new datasets and challenges.

To evaluate task 4 submissions, the primary metric used is the Polyphonic Sound Detection Score (PSDS), as proposed by Bilen et al.\cite{bilen2020framework}. This metric assesses a system's ability to correctly detect and classify overlapping sound events, with higher scores indicating better performance. In 2024, a second metric is included to complement the PSDS, the segmented-based partial area under the ROC curve (segMPAUC) \cite{cornell2024dcase}. This metric allows for more targeted evaluation in scenarios where the goal is to minimize false positives, higher values of segMPAUC also indicate better performance. The final ranking of the systems is the sum of these two metrics, the PSDS evaluated on the DESED dataset \cite{serizel2020sound,turpault2019sound}, and the segMPAUC on the MAESTRO dataset \cite{martin2023training}.

\subsection{Energy and compute metrics}
\label{sec:energycompute}
The introduction of the energy as an evaluation metric was introduced in 2022 \cite{ronchini2022description}. Participants were invited to report the energy consumption related to both the training and the testing phases of their submissions using CodeCarbon toolkit \cite{schmidt2021codecarbon}. CodeCarbon estimates the energy by tracking the power consumption of various components, including the GPU, CPU, and RAM, and sums these values to compute the total energy consumed. Since 2023, reporting energy is mandatory when participating to the task. In addition, participants are required to measure the energy consumption of the baseline system for 10 epochs of training on the same setup used for submission\cite{ronchini2024performance}. This aims to normalize energy consumption and allows for fairer comparisons between systems by taking into account potential material differences \cite{serizel2023performance}. As of this year, we ask participants to report not only the overall energy consumption, but also the component-specific energy consumption provided by CodeCarbon, especially the energy consumed by the GPU \cite{cornell2024dcase}. All energy measurements are expressed in kilowatt-hours (kWh), and the normalization factor is calculated as the ratio of the energy consumed for 10 epochs of the baseline on our setup (NVIDIA A100 40GB) to that consumed on the participants' setup. In 2024, we record 0.029 kWh for 10 epochs, compared to 0.032 kWh in 2023. Based on this normalization, we also compute the energy-weighted (EW) performance score as proposed by Ronchini and Serizel \cite{ronchini2024performance}.

In addition, we also require participants to report hardware-agnostic metrics to account for the computational load, such as the number of parameters of the system, and since 2023, the Multiply-Accumulate Operations (MACs) needed to process 10 seconds of audio. The MACs are computed using the THOP: PyTorch-OpCounter tool \cite{zhu2019thop}.

\subsection{Submissions}

Submissions are divided into single systems (or non-ensemble systems) and ensemble systems, that combines multiple systems outputs for the final evaluation. This year, each team is allowed to submit a maximum of four systems, compared to eight systems in the last edition. This change had a significant impact on the number of submissions, which dropped from 101 in 2022 and 84 in 2023 to 42 in 2024 - a reduction by 50\%. Moreover, each team is required to submit at least one non-ensemble system to limit the growing use of resource-intensive ensembles and to better understand the performance when participants are restricted to a single system. After filtering out submissions with missing or incorrect energy and compute data reports, the final number of valid submissions is 60, 64, and 35 for 2022, 2023, and 2024, respectively. The amount of non-ensemble systems is about half of the number of submissions.

\section{Energy and complexity evolution}

In this section, we analyze trends in computational cost and energy consumption of the 2022, 2023, and 2024 DCASE task 4 submissions, including normalized energy comparisons and GPU-specific energy consumption.

\subsection{General comparisons (2022 - 2024)}
\label{sec:general}

\begin{table}[htbp]
\centering
\caption{Comparison of system complexity and MACs of DCASE 2022, 2023 and 2024 submissions.}
\begin{tabular}{@{\extracolsep{\fill}}l|ccc|ccc}
 & \multicolumn{3}{c|}{\textbf{System complexity} $\downarrow$} & \multicolumn{3}{c}{\textbf{MACs} $\downarrow$} \\
Year & 25\% & Median & 75\% & 25\% & Median & 75\% \\
\hline
2022 & \SI{2.2}{M} & \SI{6.68}{M} & \SI{18.9}{M} & - & - & - \\
2023 & \SI{4.80}{M} & \SI{14.66}{M} & \SI{97.18}{M} & \SI{3.50}{G} & \SI{9.74}{G} & \SI{124}{G} \\
2024 & \SI{3.44}{M} & \SI{17.40}{M} & \SI{207}{M} & \SI{1.74}{G} & \SI{20.82}{G} & \SI{108}{G} \\
\hline
\end{tabular}
\label{tab:complexity_macs}
\end{table}

\begin{table}[htbp]
\centering
\caption{Comparison of energy consumption for training and testing of DCASE 2022, 2023, and 2024 submissions.}
\begin{tabular}{@{\extracolsep{\fill}}l|ccc|ccc}
 & \multicolumn{3}{c|}{\textbf{Energy train (kWh)} $\downarrow$} & \multicolumn{3}{c}{\textbf{Energy test (kWh)} $\downarrow$} \\
Year & 25\% & Median & 75\% & 25\% & Median & 75\% \\
\hline
2022 & 1.82 & 3.70 & 17.30 & 0.01 & 0.03 & 0.05 \\
2023 & 1.62 & 4.30 & 13.98 & 0.02 & 0.04 & 0.28 \\
2024 & 3.47 & 8.78 & 17.58 & 0.06 & 0.14 & 0.42 \\
\hline
\end{tabular}
\label{tab:energy_train_test}
\end{table}

Table~\ref{tab:complexity_macs} shows the 25th percentile, median, and 75th percentile values for system complexity and MACs over the years. We use quartiles and medians as measures of central tendency rather than mean and standard deviation due to the high variability of the data. As expected, system complexity continues to increase over the years, with a clear trend toward larger and more complex systems. In particular, the significant increase in the 75th percentile reflects that participants submitted much larger models this year than in the previous editions. The number of operations introduced in 2023 reinforces this upward trend, with the median also increasing from 2023 to 2024. However, in contrast to system complexity, the 75th percentile for MACs decreases slightly in 2024. This could be due to the inherent composition of the networks, where increasing the number of parameters does not necessarily lead to a proportional increase in the number of operations. We further investigate whether these trends are reflected in the energy consumption.

Table~\ref{tab:energy_train_test} shows the analysis for the energy consumption at training and test over the years. We see a strong increase in the training energy consumption from 2022 to 2024, the median value even doubling between 2023 and 2024. While there is a slight decrease between 2022 and 2023 at the 25th percentile, this effect is reversed and amplified in 2024, with the most efficient models consuming more energy. Similarly, at test, the energy consumption increases significantly for both the median and the second percentile, showing that more complex architectures are not only more energy intensive during training, but also at test. However, it should be noted that the energy values are not normalized and this analysis only shows general trends in energy consumption and cannot be conclusive. Therefore, we continue our analysis on the normalized energy consumption in the remainder of this study, focusing solely on 2023 and 2024 entries due to the availability of data.

\begin{figure*}[htbp]
    \centering
    \begin{subfigure}[t]{0.45\textwidth}
        \centering
        \includegraphics[width=\linewidth]{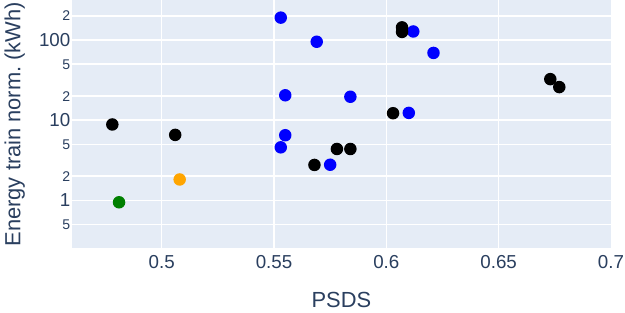}
        \caption{Ensemble}
    \end{subfigure}%
    ~ 
    \begin{subfigure}[t]{0.45\textwidth}
        \centering
        \includegraphics[width=\linewidth]{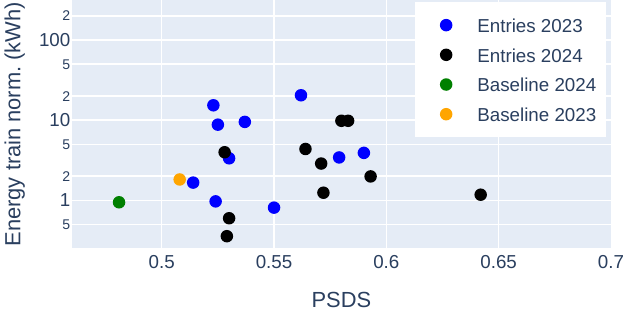}
        \caption{Non-Ensemble}
    \end{subfigure}
    \caption{Relationship between PSDS and training energy consumption for the best ensemble and non-ensemble systems from 2023 and 2024. Baselines from both years are reported as references.}
    \label{fig:energy_train_psds}
\end{figure*}

\subsection{Energy normalized comparisons (2023 - 2024)}

Table~\ref{tab:energy_norm} provides a summary of normalized energy consumption for 2023 and 2024 submissions. Note that only 60 and 30 systems in 2023 and 2024, respectively, accurately report their baseline energy consumption and are included in the remainder of the analysis. Contrary to the trends observed previously, the normalized training energy does not increase from 2023 to 2024. In fact, both of the median and the two percentiles actually decrease, reflecting a positive shift toward more energy-efficient models. The gap between the absolute energy consumption and the normalized energy is likely attributable to variations in hardware setups used by participants.

\begin{table}[t]
\centering
\caption{Comparison of energy consumption normalized for training and testing DCASE 2023 and 2024 submissions.}
\begin{tabular}{l|ccc|ccc}
 & \multicolumn{3}{c|}{\textbf{Energy train norm. (kWh)} $\downarrow$} & \multicolumn{3}{c}{\textbf{Energy test norm. (kWh) }$\downarrow$} \\
Year & 25\% & Median & 75\% & 25\% & Median & 75\% \\
\hline
2023 & 2.51 & 6.59 & 13.14 & 0.02 & 0.05 & 0.37 \\
2024 & 1.86 & 4.37 & 9.59 & 0.03 & 0.07 & 0.16 \\
\hline
\end{tabular}
\label{tab:energy_norm}
\end{table}

\begin{table}[htbp]
\centering
\caption{Comparison of the energy consumed for training 10 epochs of the baselines system in 2023 and 2024.}
\begin{tabular}{l|c}
Year &  Energy 10 epochs (kWh) \\
\hline
2023 &  0.030~$\pm$~0.020 \\
2024 & 0.052~$\pm$~0.023
\end{tabular}
\label{tab:energy_baseline}
\end{table}

In order to verify this hypothesis we compare the energy consumption reported by participants to perform 10 epochs of training for the baseline. Despite the changes between 2023 and 2024, the energy to train the baseline remained similar on our setup (0.032 kWh in 2023 vs.  0.029 kWh in 2024, see also Section \ref{sec:energycompute}) so changes in the participants reports are likely due to changes in hardware usage. In Table \ref{tab:energy_baseline}, we report the average consumption of submissions to compute 10 epochs of the baseline. As expected, there is a significant increase in energy consumption of 0.012~kWh from 2023 to 2024. These results indicate that, on average, the hardware used by participants in 2024 consumed more energy than in 2023.

In contrast, for the normalized energy consumption at test, the median and the 25th percentile increase between 2023 and 2024, in line with the system complexity and the absolute energy consumption at test. However, the 75th percentile of the energy consumption at test decreases by a factor of two, consistently to the reduction previously observed for MACs-intensive systems. It is important to note that most of these systems are designed to run on target devices with specifications far from those of a GPU (e.g., embedded devices). However, the energy consumption is measured here on a GPU where the GPU may be underutilized. To confirm the latter, we continue our analysis on the relationship between the total and the GPU-specific energy consumption.

\subsection{GPU energy consumption (2024)}

This year, we ask participant to record the GPU consumption for training and testing their systems \cite{cornell2024dcase}. We present in Table~\ref{tab:gpu_energy} the proportion of the energy consumed by the GPU relative to the total energy consumption, expressed as percentage (\%GPU). On average, we observe that the GPU accounts for half of the energy consumed during training, but this drops to 38\% at test. Moreover, the larger standard deviation and wider range of GPU usage at test reflect significant differences in the participants' hardware usage between train and test. This variability is partially due to the high utilization of the GPU during training compared to its underuse at test. 

\begin{table}[ht]
\centering
\caption{Ratio of GPU energy to total energy (\%GPU) for training and testing DCASE 2024 systems.}
\begin{tabular}{l|ccc}
&  Mean $\pm$ Std & Min & Max \\
\hline
\%GPU Train &  53\% $\pm$ 17\% & 27\% & 87\% \\
\%GPU Test & 38\% $\pm$ 21\% & 5\% & 85\% \\
\end{tabular}
\label{tab:gpu_energy}
\end{table}

Additionally, measuring the energy consumption at test on GPU may not accurately reflect the intended deployment hardware for these type of relatively small models. Therefore, for the remainder of our analysis, we focus only on training energy consumption.

\begin{figure*}[t!]
    \centering
    \begin{subfigure}[t]{0.45\textwidth}
        \centering
        \includegraphics[width=\linewidth]{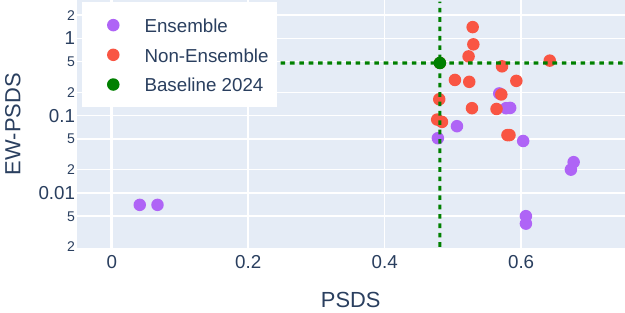}
    \end{subfigure}%
    ~ 
    \begin{subfigure}[t]{0.45\textwidth}
        \centering
        \includegraphics[width=\linewidth]{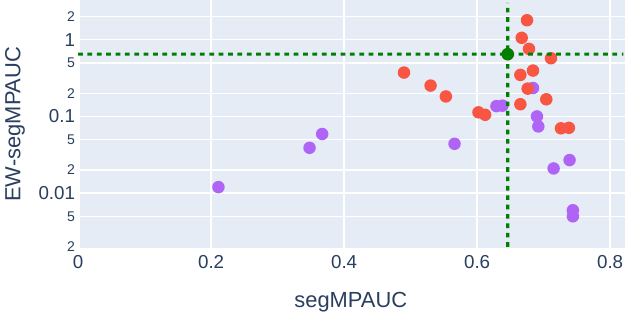}
    \end{subfigure}
    \caption{Relationship between energy-weighted (EW) metrics and performance (PSDS and segMPAUC) for ensemble and non-ensemble systems in 2024, compared to the baseline system.}
     \label{fig:energy_weighted}
\end{figure*}

\section{Performance analysis}

In this section, we put a highlight on the performance of DCASE task 4 submissions and how it relates to energy consumption. We compare ensemble and non-ensemble systems, explore energy-weighted metrics, and consider the effect of energy consumption thresholds.

\begin{table*}[ht]
\centering
\caption{Performance comparison of DCASE 2023 and 2024 non-ensemble submitted systems when thresholding system complexity, MACs, and normalized training energy consumption. For each threshold (25\% and median), the best systems in terms of PSDS below the threshold are reported. "All" represents results without any threshold applied.}
\begin{tabular}{l|cc|cc||cc|cc||cc|cc}
& \multicolumn{4}{c||}{\textbf{System complexity}} & \multicolumn{4}{c||}{\textbf{MACs}} & \multicolumn{4}{c}{\textbf{Energy train norm. (kWh)}} \\
& \multicolumn{2}{c|}{2023} & \multicolumn{2}{c||}{2024} &  \multicolumn{2}{c|}{2023} & \multicolumn{2}{c||}{2024} &  \multicolumn{2}{c|}{2023} & \multicolumn{2}{c}{2024} \\
& Max $\downarrow$ & PSDS $\uparrow$ & Max $\downarrow$ & PSDS $\uparrow$ & Max $\downarrow$& PSDS $\uparrow$ & Max $\downarrow$ & PSDS $\uparrow$ & Max $\downarrow$ & PSDS $\uparrow$ & Max $\downarrow$ & PSDS $\uparrow$ \\
\hline
All & 1G & 0.59 & 181M & 0.64 & 460G & 0.59 & 45G & 0.64 & 23.01 & 0.59 & 9.84 & 0.64 \\
25\% & 5M & 0.55 & 1.6M & 0.52 & 912M & 0.55 & 1.2G & 0.57 &0.99 & 0.55 & 1.18 & 0.53 \\
Median & 6M & 0.59 & 3.4M & 0.59 & 4G & 0.55 & 1.7G & 0.59 & 2.33 & 0.56 & 1.99 & 0.64 \\

\hline

\end{tabular}
\label{tab:threshold}
\end{table*}

\subsection{Energy and performance comparisons (2023 - 2024)}

To better understand the evolution of the normalized energy consumption at training and the relation with performance, we separate ensemble from non-ensemble systems. In total, there are 33 and 13 ensemble systems, and 27 and 17 single systems for 2023 and 2024, respectively. For this analysis, we focus specifically on the top 10 systems in terms of PSDS performance. Figure~\ref{fig:energy_train_psds} illustrates the relationship between PSDS and normalized training energy consumption. For ensemble systems, we see an increase in PSDS scores between 2023 and 2024, while the energy consumption remain in a similar range. This points out that performance improvements are achieved without a proportional increase in energy consumption from one year to another. For non-ensemble systems, we observe that it is even possible to reduce energy consumption while also increasing performance, which is a promising sign of optimization and efficiency. When comparing the two graphs in parallel, we see that all ensemble systems consume more energy than non-ensemble systems - the logarithmic scale underrepresenting the gap in energy consumption. Despite achieving better PSDS results, the significant energy cost associated with ensemble methods raises questions about their overall effectiveness. For example, in 2024, the best ensemble system achieves a PSDS score of 0.68 and consumes 25.9~kWh. In contrast, the top non-ensemble system reaches a PSDS score of 0.64 with a much lower energy consumption of 1.2~kWh. This non-ensemble system also outperforms the best ensemble system from 2023, which had a PSDS score of 0.62 but consumed 69.1~kWh.

\subsection{Energy-weighted performance (2024)}

We continue our analysis on the trade-offs between energy consumption and performance improvements by considering the energy-weighted metrics. Due to space constraint we focus on the 2024 systems only. Figure~\ref{fig:energy_weighted} shows the relationships between PSDS and EW-PSDS on the left, and segMPAUC and EW-segMPAUC on the right. The green dotted lines segment the graphs into areas where systems either consume more energy while being outperformed by the baseline (bottom-left), outperform the baseline but with higher energy use (bottom-right), or achieve better performance with lower energy consumption (top-right). Unfortunately, we observe that only a few systems are located in that optimal corner (top-right) independently of the metric, and these are exclusively non-ensemble systems. Ensemble systems are particularly concentrated towards the lower end of the bottoms quadrants, highlighting their relatively high energy costs compared to performance improvements. For the PSDS plot (left), most systems are located in the bottom-right corner, while the segMPAUC plot (right) shows a more dispersed distribution, with systems spread across the bottom-left and bottom-right quadrants.  This dispersion suggests that some systems have optimized for PSDS at the expense of segMPAUC, where the majority of the energy consumption is concentrated on this optimization. 

\subsection{Threshold based on energy consumption (2023 - 2024)}

The previous analysis shows that large energy costs can result in small performance gains, which leads us to consider the notion of an “energy cap”, simulating a scenario where participants have limited resources or budget. This approach was previously proposed by Ronchini and Serizel \cite{ronchini2024performance}, but here we extend the analysis to compare differences between 2023 and 2024. All results are presented on Table \ref{tab:threshold}. Generally, we see that setting a cap on complexity and energy metrics results in reduced performance, but the extent of the drop depends on the specific threshold level and the year. 

When looking at median energy consumption in 2024, there are notable improvements in both energy efficiency and system performance compared to 2023, which is encouraging. However, when focusing on low-energy systems (25\%), the results are less positive: in 2024, these systems perform worse and consume more energy than in 2023. This suggests a reduction in the number of high-performing, low-energy systems in 2024. A similar trend is observed in the MACs metric, although the 25th percentile shows a slight improvement despite higher MACs in 2024. 
In contrast, system complexity has decreased in 2024 (at all, median, and 25th percentile levels). While performance is improved when considering the most complex systems, it is either been maintained or deteriorated when considering only lower complexity system. Furthermore, applying the median energy cap in 2024 allows best performance to be maintained, a result not observed in 2023. This suggests that, unlike in 2023, the best performing systems in 2024 are not necessarily those with the highest training energy consumption.\footnote{Similar results for segMPAUC can be found in the additional results page.}

\section{Conclusions}

In conclusion, our analysis of DCASE Task 4 submissions from 2022 to 2024 highlights considerable evolution in system performance and energy efficiency. Despite a general trend toward increasingly complex and energy-intensive systems, we observe notable improvements in energy efficiency when accounting for hardware disparities. Specifically, the normalized energy consumption for training has decreased in 2024 while the systems performed better in terms of PSDS. However, ensemble systems still consume large amounts of energy relative to their performance gains, and few single systems outperform the baseline with lower energy use. In addition, we find that setting a cap on resources can lead to performance loss, but also highlights advances in energy efficiency. This also emphasize the energy cost for performance improvements that are sometimes relatively modest. Finally, in 2024, the best performing systems are not necessarily those with the highest energy consumption, indicating a positive shift towards more sustainable and efficient solutions.

\bibliographystyle{IEEEtran}

\bibliography{refs}

\end{document}